# Extraction of Weak Surface Diaphragmatic Electromyogram Using Modified Progressive FastICA Peel-Off

Yao Li#, Dongsheng Zhao#, Haowen Zhao, Min Shao*, Xu Zhang*, *Member, IEEE*

*Abstract*—Diaphragmatic electromyogram (EMGdi) contains crucial information about human respiration therefore can be used to monitor respiratory condition. Although it is practical to record EMGdi noninvasively and conveniently by placing surface electrodes over chest skin, extraction of such weak surface EMGdi (sEMGdi) from great noisy environment is a challenging task, limiting its clinical use compared with esophageal EMGdi. In this paper, a novel method is presented for extracting weak sEMGdi signal from high-noise environment based on fast independent component analysis (FastICA), constrained FastICA and a peel-off strategy. It is truly a modified version of of progressive FastICA peel-off (PFP) framework, where the constrained FastICA helps to extract and refine respiration-related sEMGdi signals, while the peel-off strategy ensures the complete extraction of weaker sEMGdi components. The method was validated using both synthetic and clinical signals. It was demonstrated that our method was able to extract clean sEMGdi signals efficiently with little distortion. It outperformed state-of-the-art comparison methods in terms of sufficiently high SIR and CORR at all noise levels when tested on synthetic data, while also achieved an accuracy of 95.06% and a F2-score of 96.73% for breath identification on clinical data. The study presents a valuable solution for noninvasive extraction of sEMGdi signals, providing a convenient and valuable way of ventilator synchrony with a significant potential in aiding respiratory rehabilitation and health.

## I. INTRODUCTION

DIAPHRAGM is a vital auxiliary organ for human breathing, and diaphragmatic electromyogram (EMGdi) carries important information about activity and health condition of human respiratory [1]. Therefore the EMGdi is commonly used to monitor respiratory conditions [2]. Compared with inspiratory flow and inspiratory pressure, the neural electrophysiological approach through EMGdi is a natural and effective way to assess respiratory functionality [3]. By further processing EMGdi we can get diaphragm electrical activity (EAdi), which plays a significant role in triggering and adjusting the ventilator to cooperate with human breathing [4]. Recently, an advanced type of ventilator termed neurally adjusted ventilatory assist (NAVA) [5] has been developed to control the timing and level of ventilator pressurization according to EAdi.

There are two main ways to collect EMGdi. One way is to obtain EMGdi from esophageal using a nasogastric catheter with electrodes [6]. There are certain drawbacks associated with collecting EMGdi signals from esophageal. It can cause discomfort for patients and potentially result in adverse effects such as bleeding and vomiting, leading to operational and applicable limitations [7]. The other way of obtaining EMGdi signals is through surface electrodes [8]. Surface electrodes are always positioned on the skin of subcostal region beneath the thorax, in the closest proximity to the diaphragm. Therefore surface EMGdi (sEMGdi) are often contaminated by a variety of interferences and noises, such as electrocardiogram (ECG) interference, electrode motion artifacts and environmental noises, leading to a higher demand for noise reduction [9]. Moreover, when endogenously generated EMGdi signals propagates from diaphragm to the skin, they need to pass through a series of body tissues, resulting in an extremely weak amplitude of the sEMGdi signals. Due to its low signal quality, it is challenging to extract reliable EAdi-derived biomarkers from sEMGdi recordings [10]. With these regards, how to extract respiration-associated component from great noisy recordings becomes a key technique for ensuring applicability of the sEMGdi while maintaining its critical advantage in noninvasive measurement. Currently available techniques failed yet to offer satisfactory solutions [11-15]. That's why the esophageal collection of EMGdi signals remains to be the dominant way used in NAVA ventilators on the market, despite its limitations.

Many efforts have been made for surface EMGdi denoising to extract desired respiratory-related EMGdi signals from significant noise condition. One of the most simple and cost-efficient solution is high-pass filter. However, it should be noted that the strong ECG interference is mainly concentrated in the 1 to 50 Hz frequency range, where the dominant power spectrum of EMGdi ranges from 20 Hz to 200 Hz. Therefore, the overlapping frequency of ECG and EMGdi signals makes it not effective to remove the ECG signals without causing significant loss of the EMGdi signals relying solely on traditional high-pass filters or bandpass filters [16]. Torres et al.

This work was supported in part by the Anhui Provincial Key Research and Development Project under Grant 2022k07020002. (*Corresponding authors: Min Shao and Xu Zhang.*)

# The two authors contribute equally to this work.

Y. Li, H. Zhao and X. Zhang are with the School of Microelectronics at University of Science and Technology of China, Hefei, Anhui, China. (xuzhang90@ustc.edu.cn).

D. Zhao, M. Shao are with the Intensive Care Unit of the First Affiliated Hospital of Anhui Medical University, Hefei, Anhui, China.

[11] used a method combined with least mean square (LMS) algorithm and adaptive liner combiner to cancel ECG in EMGdi signals, while it requires extra ECG signals as reference signals. Jonkman et al. [12] proposed an estimated ECG subtraction method on single channel data, estimated the ECG waveform using threshold determined prior information and subtracting it from the original signal. Many researchers used wavelet-based methods such as wavelet threshold based method [13] and wavelet-based adaptive filter [14]. These methods in the time-frequency domain are all suitable for single-channel data. Furthermore, they all require a certain level of prior knowledge: for instance, adaptive filtering-based methods rely on reference signals, and wavelet-based methods necessitate the preselection of specific wavelet functions. As a result, their processing effectiveness is heavily influenced by the morphology of the signal-channel signal.

The blind source separation (BSS) technique becomes an alternative solution for extracting desired components from mixed multi-channel signals, due to the fact that electrophysiological signals can be viewed as a mixture of various source signals within the human body. Alty et al. [15] applied a fast independent component analysis (FastICA) algorithm on four-channel EMGdi signals to remove ECG interference. However, the removal was not entirely complete even on esophageal EMGdi with better quality, resulting in unsatisfactory EMGdi extraction performance. It was because the FastICA could just provide a preliminary separation of the signals. Facing with sEMGdi signals with high noise level, the algorithm tends to converge to the larger noise components like ECG interferences. [17] also points out that the signal amplitude seems to be weaken when using FastICA for ECG cancellation. Using the traditional FastICA alone cannot guarantee accurate extraction of the desired source signals. Additionally, considering the filtering effect of body tissues such as skin and fat during the propagation process, as well as the propagation delay required for electrophysiological signals to reach different electrode sites, the obtained source signal waveforms are distinct in each electrode channel [18]. Traditional BSS methods treat the signals as instantaneously mixed models without considering the time differences in signal propagation at each moment [19], leading to further compromised signal extraction performance. Other efforts have been made by incorporating the FastICA approach with additional signal processing algorithms like wavelet analysis to remove ECG artifact [20]. Many studies just introduce some preprocessing approaches on individual signal channels before utilizing the traditional FastICA. They do not fundamentally address the above issues due to lack of sufficient understanding and precise description of the concerned signal sources. Generally, these methods are prone to local convergence, making the algorithms fail to converging to the desired components. Even if they are able to make somehow source separation, it represents only an initial separation, yielding inaccurate and incomplete results.

In recent years, Chen and colleagues [21] have reported a new FastICA-derived method called progressive FastICA peel-off (PFP). The PFP serves as an algorithmic framework incorporated with necessary constraints describing electrophysiological signal sources [22, 23]. Therefore, it was regarded to address many issues mentioned above with successful applications in the field of surface EMG decomposition [21, 24, 25]. It was reported to be capable of refining more accurate motor unit (MU) spikes through constrained FastICA approach from an initially separated spike train. In addition, it employed a peel-off strategy to avoid the local convergence to larger MU spikes. Inspired by these ideas, we believe that many issues we encounter in extracting sEMGdi signals can be addressed accordingly.

This paper presents a method based on traditional PFP framework with innovative modifications designed for extracting sEMGdi sources. Unlike any traditional PFP method for decomposing general sEMG signal into its constituent MU activities, our proposed method in this study is aimed at extracting the desired EMGdi source from highly noisy surface electrophysiological recordings. In our method, the constrained FastICA approach is specifically updated according to physiological characteristics of potential sources to refine both weak sEMGdi signal and strong ECG interference. The peel-off strategy is used as well to ensure that weak sEMGdi components can be extracted exactly and entirely. This study exhibits our efforts towards noninvasive measurement and instrumentation of weak and useful sEMGdi signals. It holds significant importance in natural assistance and precise assessment for advanced respiratory health.

## II. METHOD

### A. Data Description

#### I). Clinical Data Recording Experiments

Ten subjects without consciousness but having spontaneously breathing capability were recruited from intensive care unit (ICU) of the First Affiliated Hospital of Anhui Medical University (FAH-AMU, Hefei 230011, China). In addition, 10 healthy adults with full consciousness were also recruited from the University of Science and Technology of China (USTC, Hefei 230026, China) to participate the data collection experiments. This study was approved by the Institutional Ethics Committee of the FAH-AMU under Application No. PJ 2024-05-84, and the Biomedical Ethics Review Broad of USTC under Application No. 2024KY188. Before any procedure of the experiment, informed and signed consent was obtained from the subjects or their conservators.

Multi-channel surface EMGdi signals were recorded from the skin at participant's right costal arches, specifically between the 8th and 10th rib cartilage [26], using a flexible 32-channel mono-polar electrode array arranged in an $4 \times 8$ grid as Figure 1(a) shows. Each electrode of the array was in a diameter of 2 mm with an inter-electrode distance of 4 mm. A multichannel surface electromyogram data recording system (FlexMatrix Inc., Shanghai, China) was used for data recording. It was built with a two-stage amplifier at a total gain of 60 dB, a band-pass filter set at 1-500Hz for each channel and an analog-to-digital converter (RHD 2132, Intan technologies), with a sampling rate of 1 kHz.

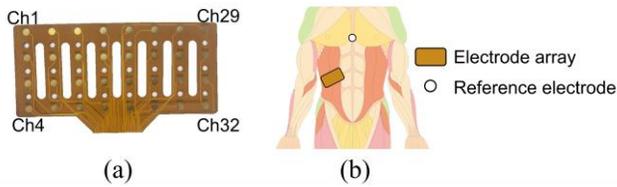

Figure 1. (a) The multi-channel electrode array used in our study. (b) Illustration of the electrode position

After the skin preparation using medical alcohol, the electrode array was placed over the skin at patient's right costal arch. An Ag-AgCl-based self-adhesive surface electrode was placed superior to the tip of the xiphoid process as the reference, which is shown in Figure 1(b). In each trial, unconscious but spontaneously breathing subjects laid supine on the bed with ventilator assistance. Their sEMGdi signals were measured for time durations ranging from 15 seconds to 2 minutes, while conscious subjects were asked to breathe at varying intensities and speeds. Real-time expiratory airflow was measured with a gas flowmeter during experiment as the ground truth.

The recorded raw sEMGdi data were generally pre-processed to reduce the noise contamination. A Butterworth bandpass filter set at 20-150 Hz was applied to eliminate the potential low-frequency motion artifacts and high-frequency environmental interferences. Subsequently, a set of notch filters were utilized to remove power line interference as well as its harmonics.

*II）. Synthetic Data Simulation*

The acquired raw signal was considered as a mixture of pure EMGdi signal, ECG signal, and various unavoidable noises recorded at each surface electrode [27]. Based on the assumption that the recorded waveform and amplitude of each signal source may vary across electrodes placed in different areas, a shift-invariant convolution model is introduced to describe the raw signals recorded by multiple channels [28, 29]. The synthetic data used in this paper is composed by three parts:

1. Clean EMGdi signals. A model-based approach was conducted for simulating multi-channel sEMGdi signals of the considered muscle [30]. According to the experiments, we also simulated sEMGdi signals recorded by a 32-channel surface electrode array in a $4 \times 8$ grid form, which is conformity to clinical scenarios, the amplitude and waveform of the signals varied across each channel.
2. Pure ECG signals. We placed the same electrode on the left chest of the subject who was asked to maintain completely relaxed, to record signals without influence of any voluntary or spontaneous muscle contraction. These signals were considered as pure ECG signals.
3. Gaussian noise. Noise besides ECG interference are considered to follow Gaussian distribution. Therefore we used a period of Gaussian noise to represent them in the synthetic data.

By directly mixing each trial of clean EMGdi signals, pure ECG signals and Gaussian noises based on the formula:
$$X_{syn} = EMGdi_{clean} + b \times ECG_{pure} + b^2 \times noise \quad (1)$$
we obtained 32-channel synthetic contaminated EMGdi signals $X_{syn}$, while $ECG_{pure}$ denotes the pure ECG signals, $noise$ denotes the gaussian noise, $b$ represents the mixing coefficient of ECG and gaussian noise to obtain synthetic data with different noise level. To investigate the performance of proposed method on signals with different noise level, synthetic signals with SIR value of -20dB, -10dB, 0dB and 10 dB were obtained respectively by setting $b$ to 4.6, 1.5, 0.5, 0.15 according to (1). Figure 2 shows a heatmap corresponding to the 32-channel electrode array used in the study with two representative signals with different amplitudes, representing the spatial amplitude of the synthetic signals. It can be observed that the signals gradually weaken as they propagate from left to right, which is consistent with the data collected in clinical settings.

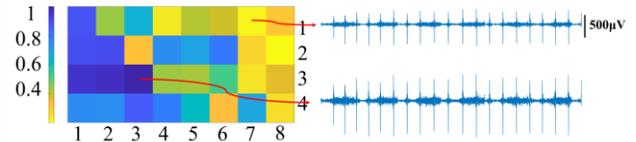

Figure 2. Heatmap of signal amplitudes corresponding to the electrode used in this study, arranged in a 4×8 grid, with two representative signals with different amplitudes.

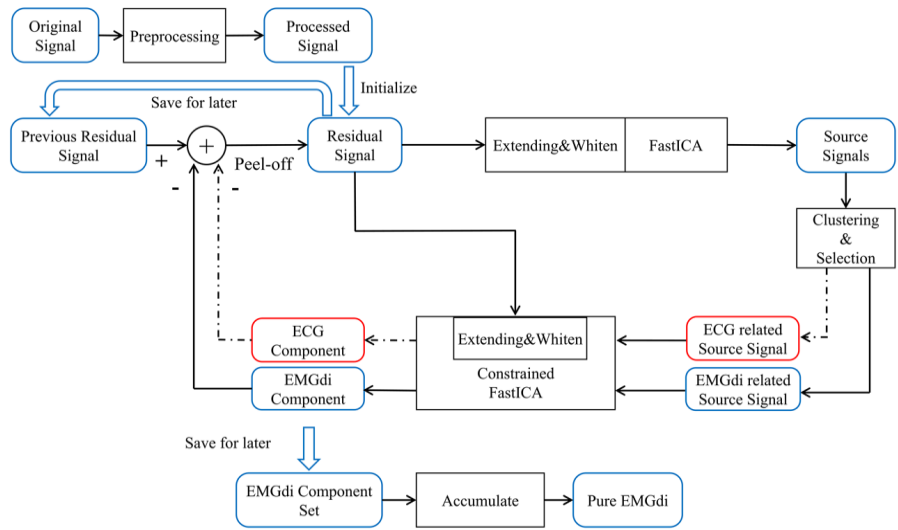

Figure 3. The flowchart of the signal processing framework

*B. sEMGdi Extraction Framework*

Figure 3 illustrates the flowchart of the framework used in this study, which includes a preprocessing approach, a FastICA followed by a constrained FastICA, a peel-off strategy and necessary component identification and selection modules. The raw signals collected is subjected to 20Hz to 150Hz band-pass filtering, 50Hz and harmonic band-stop filtering, finally acquire the preprocessed signal. The processed signal is then feed into FastICA module for preliminary source separation. Based on the different characteristics among these source signals, the initial source signals obtained, including those related to ECG and EMGdi, are selected and fed into constrained FastICA for further separation to obtain more accurate components. With the usage of constraint conditions to make the source separation process more accurate and complete, the output of constrained FastICA can result in a closer approximation to the true source signals [21]. Detailed descriptions are as follows.

*I). ECG and EMGdi source separation*

Assume that the signal $X = [x_1, x_2, ..., x_m]^T$ we collected and preprocessed can estimate the source signal using a demixing matrix $y = w^T x$ base on the blind source separation concept. For the extended and whitened residual signals, the following optimization problem needs to be solved [24]:

$$max \quad J_G(w) = [E\{G(w^T x)\} - E\{G(v)\}]^2$$
$$s.t. \quad h(w) = E\{y^2\} - 1 = \|w\|_2^2 - 1 = 0 \quad (2)$$

where $v$ is a standard normal random variable, $G$ is a nonquadratic function where we can use $G(x) = \log(\cosh(x))$ as a default. We can iterate to update according to the formula [24]:

$$w^+ = E\{xG'(w^T x)\} - E\{G''(w^T x)\}w$$
$$w = w^+/\|w^+\|_2 \quad (3)$$

until $|w_{k+1} - w_k| < \theta$, where $\theta$ is the convergence threshold.

In order to facilitate the application of FastICA and improve the numerical conditioning, previously discussed data model can be extended in the channel direction by $K-1$ delayed repetitions of each observation [18]:

$$\bar{X} = \begin{matrix}[X_1(t), ..., X_1(t-K+1), \\ ..., X_M(t), ..., X_M(t-K+1)]^T\end{matrix} \quad (4)$$

where the delay factor K denotes the total number of time intervals to be delayed. Considering this trade-off, the delay factor K was determined to be 5 in this paper. After applying FastICA on $\bar{X}$, a rough estimate of ECG and EMGdi source signals can be separated.

*II). EMGdi and ECG source selection*

The FastICA-derived output source signals are categorized into different clusters. For each cluster, its relevance with prior respiratory and ECG signals was calculated using a dynamic time warping algorithm. On this basis, whether each cluster represents EMGdi, ECG or noise source can be determined.

*III). Refining source signals using Constrained FastICA*

The FastICA outputs may just be preliminarily separated source signals, especially when some source signals have greatly diverse amplitudes. To further assess and validate the extracted source components, a constrained FastICA is used in our study by adding some constraint functions related to the expected signal. Therefore, more accurate source signals can be obtained from the initial FastICA processing [31]. Compared to the FastICA, the optimization problem of the constrained FastICA is described below. For the preprocessed signal $X$,

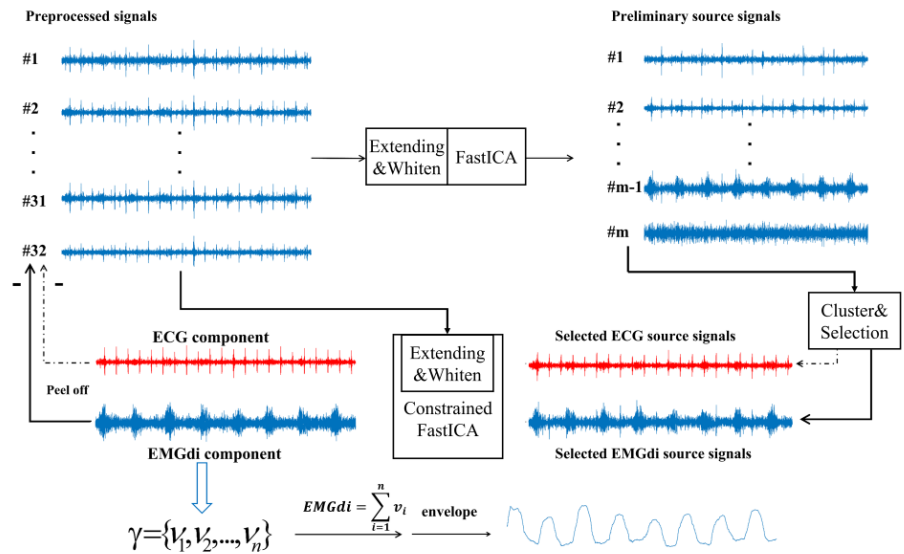

Figure 4. Illustration of the sEMGdi extraction using the proposed method

$$\begin{aligned} max \quad & J_G(w) = [E\{G(w^T x)\} - E\{G(v)\}]^2 \\ s.t. \quad & h(w) = E\{y^2\} - 1 = \|w\|_2^2 - 1 = 0 \\ & h(w) = E\{(w^T x)^2\} - 1 = \|w\|^2 - 1 = 0 \\ & E\{r^2\} - 1 = 0 \end{aligned} \quad (5)$$

where $g(y)$ measures the similarity between the output $y = w^T x$ and the reference signal $r$. In order to ensure the convergence of the algorithm, the time-domain constraint threshold can be initially set to a large value close to 1 (such as 0.99), and then gradually reduced during the iteration process. Just like FastICA without a constraint, the constrained FastICA can also be applied on the extended form of signals, with the selected ECG component or EMGdi component as reference signal, as described in Eq. (4). As compared with the FastICA, its constrained version requires a relatively larger value of $K$ for improved performance. Therefore, the delay factor K used for the constrained FastICA process was set to be 15 in this study. Such a constraint is able to drive FastICA to converge toward and independent component mostly similar to the source signals.

IV). *Extracting source signals using a peel-off strategy*

The output of constrained FastICA above includes relatively accurate ECG or EMGdi source signals. However, the separation might be incomplete yet, especially after the initial run of the BSS process. It tends to obtain larger ECG interference sources. Even after multiple runs of the BSS process, it tends to provide such local results of ECG interference sources. To eliminate the impact of major interferences and further detect the weak EMGdi source signals, the peel-off strategy is employed. Whenever the constrained FastICA is used to obtain ECG or respiratory-related EMGdi source signal, its waveform corresponding to each of all channels is estimated from the input signal (also residual signal from the latest peel-off round):

$$A = (Y^T Y)^{-1} Y^T X \quad (6)$$

where $X$ denotes the input/residual signal, $Y$ denotes the output of constrained FastICA (a separated, more precise EMGdi or ECG source signal), $A$ denotes the matrix which meets:

$$\min(X - A * Y)^T (X - A * Y) \quad (7)$$

Then the residual signal $X = X - YA$ is updated after the current round of peel-off operation.

Such a peel-off approach is expected to prevent the BSS algorithm from converging locally to larger ECG signals, thus facilitating weak EMGdi components to be revealed and correctly identified.

Based on the above descriptions, the pseudocodes of our FastICA-based method for extracting clean weak EMGdi from multichannel surface EMGdi signals are given in Algorithm 1. $\bar{\bar{x}} = [\overline{x_1}, ..., \overline{x_m}]^T$ denotes the residual signal, $\tilde{v} = [\widetilde{v_1}, ..., \widetilde{v_n}]^T$ denotes EMGdi component, $n$ denotes the number of EMGdi component. The illustration of the sEMGdi extraction procedure is shown in Figure 4.

*C. Performance Evaluation*

The following metrics were used for quantitative performance evaluation in this study. When using synthetic data, we calculated the signal to interference ratio (SIR) which was defined to measure the level of ECG contamination for a signal:

$$SIR(i) = 10.\lg\left(\frac{EMGdi_i^2}{(EMGdiC_i - EMGdi_i)^2}\right) \quad (8)$$

where $EMGdiC_i$ means the contaminated EMGdi, $EMGdi_i(t)$ means the pure EMGdi in $i$ th channel. It is obvious that the higher SIR for the signal, the more it is related to respiration-related EMGdi and is influenced less by ECG and noise.

The correlation coeffiecient (CORR) was also used to measure the linear correlation between two signals:

$$CORR = \frac{\sum_{i=1}^{n}(EMGdi_i - \overline{EMGdi_i})(EMGdiC_i - \overline{EMGdiC_i})}{\sqrt{\sum_{i=1}^{n}(EMGdiC_i - \overline{EMGdiC_i})^2 (EMGdi_i - \overline{EMGdi_i})^2}} \quad (9)$$

where $n$ represents the sample number, $EMGdiC_i$ is the root mean square (RMS) envelope of contaminated EMGdi signal envelope (before processing) with a sliding window or the processed EMGdi signal envelope for different situations. $\overline{EMGdi_i}$ and $\overline{EMGdiC_i}$ represents their mean value. Noted that due to potential imperfections in the reconstructed signals at a detailed level, and we only require the envelope of EMGdi in practice. Therefore, envelopes of the signals are utilized for computing correlations.

The median frequency variation ration (MFVR) was used to measure the influence of the interference on the frequency of EMGdi component which is defined as:

$$MFVR(i) = \frac{|MF_{EMGdi_i} - MF_{EMGdiC_i}|}{MF_{EMGdi_i}} \times 100\% \quad (10)$$

Where $MF_{EMGdi_i}$ is the median frequency of clean EMGdi in $i$ th channel, and $MF_{EMGdiC_i}$ denotes the median frequency of the contaminated EMGdi signal (before processing) or the filtered EMGdi signal (after processing). Here, median

---

**Algorithm 1. Framework for FastICA based EMGdi signal extraction**

Step 1. Applying high-pass, low-pass, and bandpass filtering to the original signal and serves it as the preprocessed signal.
Step 2. Initialize the signal obtained in Step 1 as initial residual signal $X$.
Step 3. **while** n>0 **do**
    1. Extending and whiten $X$ to get $\bar{\bar{X}}$.
    2. Apply FastICA on extending and whiten $\bar{\bar{X}}$ to extract a new source signal $v$.
    **if** $v$ is a EMGdi or ECG source signal **then**
        1. Apply the constrained FastICA algorithm on $\bar{\bar{X}}$ with $v$ as reference signal, obtain more precise component $\tilde{v}$.
        **if** $v$ is a EMGdi source signal **then**
            Store $\tilde{v}$ to EMGdi component set $\gamma$.
    **end if**
    2. Estimate the waveform of $\tilde{v}$ in each channel of $X$ to get $\hat{s}$. Update $X = X - \hat{s}$.
    **end if**
**end while**
Step 5. Accumulate components in $\gamma$ to get clean EMGdi.

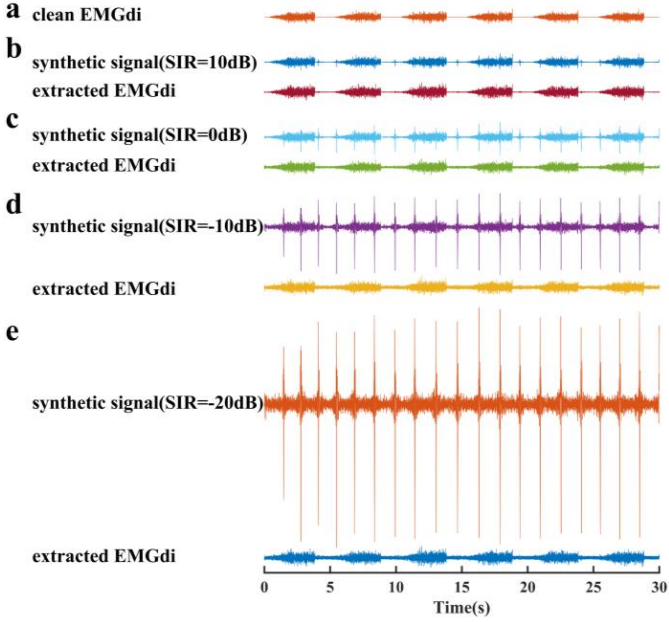

Figure 5. **a** clean EMGdi. **b** synthetic signal and extracted EMGdi with the SIR of 10dB. **c** synthetic signal and extracted EMGdi with the SIR of 0dB. **d** synthetic signal and extracted EMGdi with the SIR of -10dB. **e** synthetic signal and extracted EMGdi with the SIR of -20dB.

frequency is defined as a frequency at which 50 % of the total power of a signal segment is reached.

In order to quantitatively demonstrate the breath detection performance，recall, precision and accuracy were used, and considering that recall is more important in the extraction of respiratory signal, we also used F2-score, which are given as:

$$recall = \frac{TP}{TP + FN} \times 100\%$$
$$precision = \frac{TP}{TP + FP} \times 100\%$$
$$accuracy = \frac{TP}{TP + FP + FN} \times 100\% \quad (11)$$
$$F2 = \frac{5 \times precision \times recall}{4 \times precision + recall} \times 100\%$$

where TP denote the number of correctly detected breath, FN denote the number of missed breath and FP denote the number of incorrect breaths. Following the triggering routine of clinical ventilators, a threshold was chosen as the initial point for initiating a breath based on both the normalized actual airflow curve and the envelope extracted from sEMGdi signals. In this study, a threshold of 0.3 was employed, signifying that a breath is determined when the normalized curve reaches into 30% of the next peak. If the triggering time obtained from the airflow curve and the sEMGdi differ by less than 500ms, the breath judged by the sEMGdi was regarded to be correct.

When using clinical data, we used recall, precision, accuracy and F2-score to quantitatively demonstrate the breath detection performance, and we were not able to calculate SIR, CORR and MVFR due to the lack of ground-truth breath information.

Three representative comparison methods were also selected: the LMS adaptive filter with an extra ECG signal as reference signal; the bandpass of 50-100Hz, which excluded the main frequency range of ECG signals while preserving the main frequency range of EMGdi; the traditional blind source separation method FastICA. They were consistently operated with clinical routine or customized for optimal performance.

*D. Statistical Analysis*

In order to better examine the performance of different EMGdi extraction methods in clinical conditions, a two-way repeated-measure ANOVA was applied on the recall, precision, accuracy and F2-score, with the four methods: LMS, bandpass filter, FastICA and proposed method. If necessary, multiple pairwise comparisons with LSD corrections were performed. The level of significant difference was set as $p < 0.05$. All statistical analyses were performed using SPSS software (version 27.0, SPSS Inc. Chicago, IL, USA) in this study.

### III. RESULTS

*A. Results of Synthetic Data*

Fig. 5 exhibits representative examples of extracting clean EMGdi from synthetic raw data with different levels of noise contaminations (SIR varying from -20 dB to 10 dB for every 10 dB), where just one typical channel of the raw data is shown. By applying the proposed method, the extracted EMGdi signal is shown below its corresponding one-channel raw signal. It can be visually observed that the extracted EMGdi signal is quite consistent with the original clean EMGdi signal (Fig. 5a), regardless of the noise level.

In addition, Fig. 6 illustrates the change in CORR between the clean EMGdi and extracted EMGdi obtained by setting different peel-off numbers. Note that we just considered the peel-off number for extracting EMGdi components, where the peel-off number for detecting ECG components was not taken into account (they were not always the same). When the peel-off number was increased from 1 to 2, the CORR rose up evidently to a relatively high level, while it seldom climbed up with little fluctuations at further increased numbers. Therefore this number was set to 2 in the following analyses for optimal and stable performance. Specifically, it is interesting to find that better extraction performance was achieved at greater SIR levels when just one peel-off operation was applied.

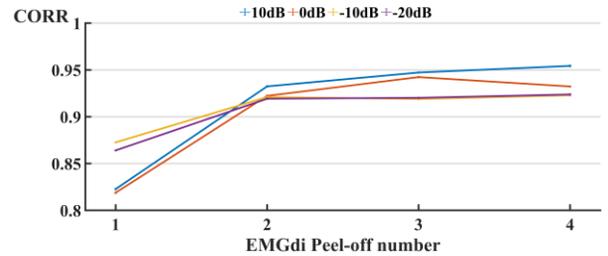

Figure 6. The variation of CORR with the number of EMGdi peel-off at four noise levels.

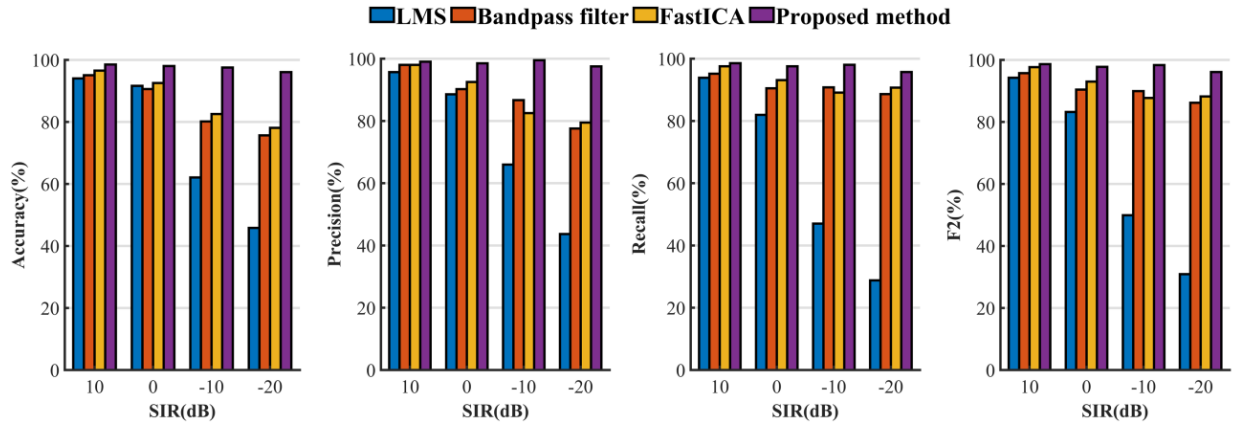
Figure 7. The performance evaluation metrics of respiratory detection on synthetic data by four different methods at four noise levels.

Table.1 shows the change of SIR after applying different methods at different noise levels. The use of the proposed method led to evident SIR improvement from its original level to a greater level. The SIR yielded by the proposed method is the highest among all methods. By contrast, such SIR improvements from other methods are very limited at any noise level, and in some cases, and they even had a SIR drop below its original level. Table 2 shows the MFVR after filtering the raw signals at different noise levels. The significantly lower MFVR (approximating to 0) further demonstrate the proposed method has little damage to the frequency of EMGdi. Table 3 presents the CORR between extracted and the original EMGdi signals. It can be seen that the proposed method achieved CORR values higher than those from other methods at any noise level, with the maximum CORR reaching into 0.9543 at an original SIR level of 10 dB. It just slightly decreased to 0.9238 at a high noise level (SIR is -20 dB). Other comparison methods failed to achieve high CORR value over 0.90, and specifically, the LMS method also encountered a severe CORR drop from 0.5803 to 0.0045 when the original SIR level was decreased from 10dB to -20dB.

Fig. 7 shows the quantitative metrics of respiratory detection on synthetic data at four noise levels using four different methods. It can be found that the performance of the LMS method declines sharply with increasing noise levels, while the bandpass filter and FastICA methods also exhibit a certain degree of performance degradation, especially in terms of precision, which drops dramatically. The proposed method achieved relatively the best performance across all noise levels. Specifically, its high performance was almost not affected by any low SIR levels.

### B. Results of Clinical Data

A series of experimentally collected data were processed by the proposed method for extracting pure sEMGdi signals. One typical example was reported in Fig. 8, where a 40-s data segment involving four slow breaths followed by seven relatively fast breaths. From visual inspection, although the preliminary source signal in Fig. 8**b** exhibits good separation of majority of strong ECG interferences, some ECG spikes are still present. The final output in Fig. 8**c** after two rounds of peel-off operations seems to carry pure sEMGdi signal without any discernable ECG interference. When its time-varying RMS envelope is calculated as the black bold curve in Fig. 8**d**, it is found to be highly correlated and satisfactorily matched with the reversed respiratory airflow curve in Fig. 8**e**. The RMS envelops of 32 individual channels of raw sEMGdi recordings are also plotted in various colors in Fig. 8**d** for comparison. They contained repetitive large fluctuations according to ECG interferences, being unsuitable for representing respiratory activities. The vertical lines in both Fig. 8**d** and Fig. 8**e** correspond to the detected triggering events when the curve reaches upon an appropriate trigger threshold. The timing of each vertical line indicates occurrence of one detected breath intention.

Table 1. Comparison of SIR of different methods at different noise levels

| Synthetic signal(dB) | LMS(dB) | Bandpass filter(dB) | FastICA(dB) | Proposed method(dB) |
|---|---|---|---|---|
| 10 | 2.6482±1.3638 | 1.4910±2.0585 | 8.9919±2.4581 | 21.7590±0.2686 |
| 0 | -3.5429±1.4392 | -2.4391±1.5401 | 0.1937±3.2144 | 10.4816±0.1692 |
| -10 | -4.1391±1.2948 | -6.2341±1.4542 | -3.0583±2.3413 | 7.1073±0.1959 |
| -20 | -8.4323±2.4492 | -12.2341±2.3129 | -5.3272±2.5943 | 6.7416±0.2173 |

Table 2. Comparison of MFVR of different methods at different noise levels

| Noise level(dB) | LMS(Hz) | Bandpass filter(Hz) | FastICA(Hz) | Proposed method(Hz) |
|---|---|---|---|---|
| 10 | 31.5934±18.3580 | 13.4672±8.3725 | 38.4943±27.5490 | 2.0782±1.3243 |
| 0 | 36.4948±15.4803 | 18.3242±7.4397 | 41.6904±21.4343 | 2.1058±1.5379 |
| -10 | 44.0354±20.6904 | 18.8594±8.4924 | 45.9010±16.4532 | 3.5930±2.4824 |
| -20 | 42.5930±13.5946 | 21.5890±9.0193 | 40.4368±25.5930 | 2.9483±1.5943 |

Table 3. Comparison of CORR of different methods at different noise levels

| Noise level(dB) | LMS | Bandpass filter | FastICA | Proposed method |
|---|---|---|---|---|
| 10 | 0.5803±0.0016 | 0.7018±0.1963 | 0.8278±0.0064 | 0.9543±0.0198 |
| 0 | 0.1955±0.0011 | 0.7100±0.1039 | 0.8064±0.0132 | 0.9323±0.0122 |
| -10 | 0.1262±0.0060 | 0.7180±0.1206 | 0.7966±0.0049 | 0.9229±0.0138 |
| -20 | 0.0045±0.0014 | 0.6897± 0.1190 | 0.7578±0.0146 | 0.9238±0.0178 |

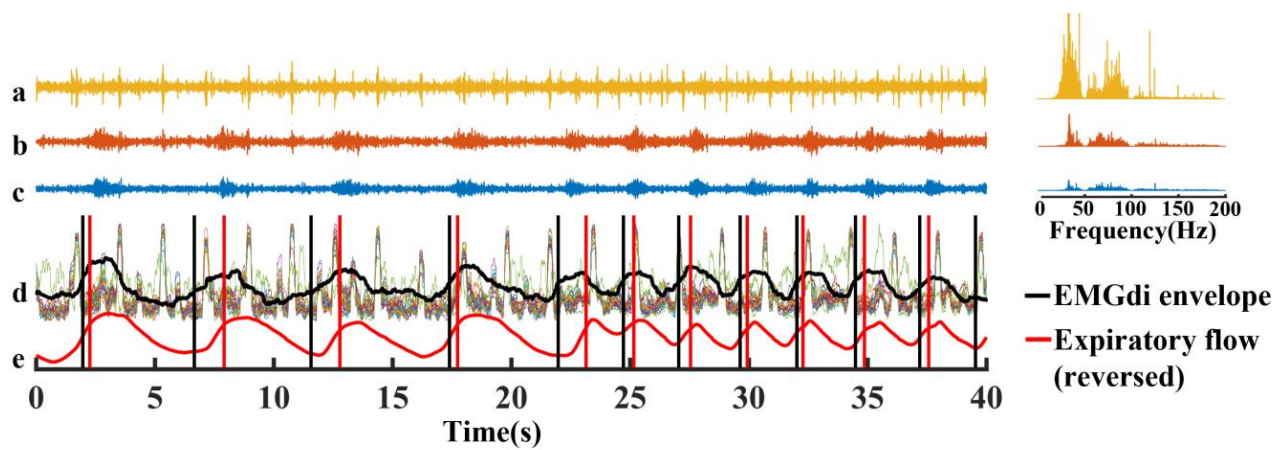

Figure 8. Illustration of extracting sEMGdi signal from a representative segment of experimentally recorded noisy sEMGdi signals in clinic. **a**. One typical channel of the raw sEMGdi signal with severe repetitive ECG. **b**. One FastICA output representing a preliminarily separated source signal corresponding to the sEMGdi, where a majority of ECG interference was separated successfully. **c**. The extracted EMGdi signal by the proposed method after 2 peel-off operations, which is the summation of two sEMGdi components. **d**. Envelope of the extracted EMGdi signal in black bold, with other 32 envelopes in different colors derived from 32 channels of the recorded noisy sEMGdi signals respectively. **e**. The reverse of corresponding airflow curve experimentally measured as true respiratory activity, which is visually found to be highly correlated to the envelop of extracted sEMGdi signal.

The respiratory detection results in terms of four metrics are shown in Fig. 9, when the proposed method and 3 comparison methods were used to process all of the clinical data. Our method incorporated with both the constrained FastICA and the peel-off strategy yielded an accuracy of 95.06±1.54%, a precision of 98.72±0.54%, a recall of 96.25±0.54% and a F-2 score of 96.73±0.87%, achieved superior performance over any comparison method in any metric, with statistical significance ($p < 0.05$).

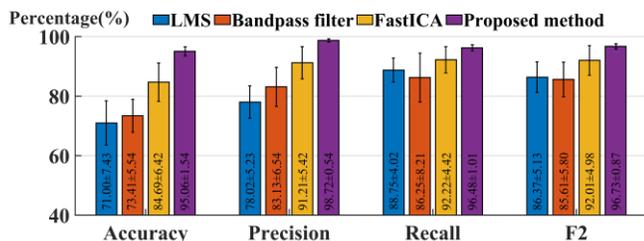

Figure 9. Respiratory detection performance in terms of four metrics when clinical data are processed by four different methods, respectively.

## IV. DISCUSSIONS

In NAVA, correct identification of pure sEMGdi signals from noise interferences is a key challenge. We developed a novel method inspired by the PFP framework for extracting weak sEMGdi signals from highly noisy recordings. One of both key designs of the method is the peel-off strategy, which can facilitate extraction of any weak source signal by removing relatively large interference sources in advance. Such a strategy helps to force the algorithm to converge to our desired weak sEMGdi signals rather than strong ECG interferences and noises. As shown in Fig. 6, extracted signals becomes more precise with the increasing of peel-off numbers. The other key design is the two-step combination of FastICA and constrained FastICA, which can be used to refine components when we extracted respiration-related signals, with the component selection module selecting the desired sEMGdi component. This can effectively improve the performance of breath detection, which is crucial in clinical application. The proposed method was validated using both synthetic signals and clinical sEMGdi signals.

For synthetic signals, as shown in Table 1, 2 and 3, our method has achieved significant effectiveness in extracting respiratory-related signals compared to other methods. It can be seen that in four noise levels, the proposed method provides the best extraction performance. Other methods without waveform reconstruction have limited improvements in SIR, and in some cases, their filtered signals may be much smaller than the original signals due to energy attenuation, therefore the SIR may even decrease. This is also consistent with a previous report [17]. Additionally, unexpected frequency variations take place as well due to the inaccurate and incomplete extraction of source signals during processing. The increase in MFVR when using traditional frequency-domain band-pass filtering method is primarily due to its inability to fully extract the required signal while filtering out noise. Information of EMGdi overlapped with the noise and artifacts is significantly attenuated, leading to pronounced spectral distortion. This highlights the limited performance of such methods. Meanwhile, a decreasing trend of CORR was found with the increasing of noise level. The degradation in extraction performance of LMS becomes particularly worse as the noise level increases. This is because it relies heavily on the inherent morphology of the reference signal, making it ineffective when the noise besides ECG also has a larger amplitude, which is common in high-noise environments. These findings are consistent with the limitation mentioned in [12]. In contrast, the proposed method has a better performance in low SNR environments, which are commonly encountered in practical applications.

The sole FastICA method achieved a slightly higher extraction performance, with higher SIR, CORR than single channel methods, however not comparable than proposed method. Traditional blind source separation algorithms, like reported in [15, 20, 32], lack of sufficient understanding and precise description about specific signals during the processing. For example, a poor performance of traditional ICA has been shown in [15], which exhibited almost no denoising effect apparently. Especially at low SIR levels, traditional FastICA tends to converge on larger ECG components than weaker EMGdi components. This local convergence makes traditional BSS methods fail to work in high-noise environment. It is also confirmed by the data reported in Table 3 which represent the performance of synthetic EMGdi signals extraction. In high-noise conditions, the FastICA algorithm often fails to converge, making it difficult to include these non-convergent cases in the statistics. As a result, the calculated CORR values of FastICA are higher than the actual situation.

Facing these issues, the proposed method using a peel-off strategy. In each round of peel-off operation, we remove the larger amplitude ECG signals and the already identified EMGdi signal components from the residual signal. As shown in Fig. 6, as the peel-off numbers of EMGdi increases, the CORR between the extracted signals and the pure EMGdi signals rise up evidently to a relatively high level, which is better than using FastICA alone. For synthetic signals used in this experiment, the signal quality significantly improved during the second round of peel-off operation, with little noticeable change thereafter. Compared to traditional blind source separation methods, our method removes the larger amplitude ECG signals and the already identified EMGdi signal components from the residual signal, therefore effectively addresses the issue of local convergence and maintains good performance even in high-noise conditions. In fact, before each peel-off procedure of an EMGdi component, there are always peel-off of ECG components because FastICA is more likely to converge to larger ECG components. It is the peel-off strategy that enables the algorithm to converge towards and extract weaker EMGdi components.

As shown in Figure 6 and Table 1, 2 and 3, the proposed method had better performance in terms of any metric than the FastICA alone, with the refining ability of the constrained FastICA module. This is because with the initial separated source signals as reference, the constrained FastICA is able to produce output signal much closer to the reference signal, resulting in less remained ECG components in EMGdi component [21]. We can also observe that although FastICA performs better than the bandpass filter in terms of the metrics presented in Table 1, 2 and 3, Fig. 7 shows that their performance in quantitatively determining respiratory intent is comparable. This is because these remaining ECG signals have minimal impact on the calculation of metrics for an entire signal segment but are crucial in determining respiratory triggers.

When dealing with clinical signals, we also tested several comparison methods for their processing effects on clinical data as shown in Fig. 9. The performances of these methods are acceptable in certain situations, such as low-noise conditions. In the discussion of synthetic signals, the proposed method has been demonstrated to be superior to other methods. Statistical metrics in Fig.9 also demonstrate the excellent performance of the proposed method in respiratory signal extraction. It is because remained ECG and noise that are not completely removed can also be mistakenly identified as a respiratory event, especially in environments with higher noise levels. Our method incorporates rich spatiotemporal information from all 32 channels, the experiments on synthetic signals have demonstrated that it achieves a good performance on the reconstruction of source signals, with constrained FastICA to refine these mistaken or missing components, thus the extracted respiratory-related signals from clinical signals have a best performance among four methods, as quantitatively demonstrated in Fig.9. The extracted source signals contain fewer other components, leading to a significant reduction in false positives and achieving substantial performance improvement. As shown in Fig.8 **b** and **c**, it is evident that there are fewer remained ECG components in Fig.8 **c** compared to **b**. Furthermore, as shown in Fig. 8 **d**, the envelope extracted by our method is superior to the envelop derived from any individual channel. Our method utilizes the spatiotemporal information from all channels, therefore can achieve the optimal solution, reducing the reliance on the placement position of individual electrodes as well as mitigate the risk of electrode positioning affecting measuring performance [33], resulting in purer and more complete extracted source signals, therefore superior to any single-channel method.

During clinical test, the envelope of the extracted sEMGdi signal aligns well with the airflow curve at different respiratory rates. Previous studies have shown that the electromyographic activity of the diaphragm directly reflects the neural commands of human respiration, while measurements of respiratory pressure or airflow through the airway exhibit certain delays. Research such as [8, 34-38] also indicates that in the future, ventilators using NAVA modes must be superior to traditional pressure or airflow triggering. As shown in Figure 8 **d** and **e**, we can also observe that the black vertical lines representing breaths triggered by sEMGdi always appear slightly earlier than the red vertical lines representing breaths triggered by airflow. Our proposed method not only achieves high accuracy in extracting respiration-related signals but also utilizes electrophysiological signals with a time advance, which offers the advantage of low latency on real-time respiration monitoring. Therefore, it holds significant importance for clinical assistance in patient respiratory rehabilitation and human-machine interaction.

The limitation of our research is that it is an offline algorithm. Therefore, the significant computational burden of the algorithm makes it impractical for real-time implementation on clinical requirement when instant respiratory curves of subjects are needed. Nevertheless, Zhao et al. [39-41] have achieved a series of innovative works in the field of online electromyographic decomposition based on the PFP method recently. This can inspire us to develop an online version of our algorithm to better meet the clinical demand for real-time output in certain scenarios.

## V. CONCLUSION

A novel method based on FastICA and peel-off strategy is presented for extracting respiration-related sEMGdi signals from high-noise original signals in this study. The proposed method uses peel-off strategy and a two-step combination of FastICA and constrained FastICA to extract high quality sEMGdi which has a superior performance of respiration identification both on synthetic and clinical sEMGdi signals. The results offer a new method to noninvasively extract weak sEMGdi using surface electrodes for triggering NAVA ventilator and monitoring respiratory status, which holds great significance for clinical human-machine synchronization.